\documentclass[a4paper,fleqn,usenatbib]{mnras}

\usepackage{mathptmx}

\usepackage[T1]{fontenc}
\usepackage{ae,aecompl}

\usepackage{graphicx}	
\usepackage{amsmath}	
\usepackage{amssymb}	
\usepackage{multicol}        
\usepackage{bm}		
\usepackage{pdflscape}	

\usepackage{epsfig}
\usepackage{threeparttable} 

\usepackage{epstopdf}
\usepackage[usenames, dvipsnames]{color}

\usepackage{booktabs,array}

\usepackage{color}

\newcommand{\SLX}{SLX~1737--282}

\newcommand{\Fx}{$F_\mathrm{X}$}
\newcommand{\lx}{$L_\mathrm{X}$}
\newcommand{\ledd}{$L_\mathrm{Edd}$}
\newcommand{\Nh}{$N_{\rm H}$}

\newcommand{\qhis}{$\chi^{2}$}

\newcommand{\lum}{\mathrm{erg~s}^{-1}}
\newcommand{\flux}{\mathrm{erg~cm}^{-2}~\mathrm{s}^{-1}}
\newcommand{\cnts}{\mathrm{counts~s}^{-1}}
\newcommand{\nh}{$\mathrm{cm}^{-2}$}

\newcommand{\xmm}{\textit{XMM-Newton}}

\newcommand{\suzaku}{\textit{Suzaku}}

\newcommand{\ergs}{erg s$^{-1}$}

\title[\SLX: Soft X-ray spectroscopy ]{The very faint hard state of the persistent neutron star X-ray binary \SLX\ near the Galactic centre}

\author[M. Armas Padilla et al.]{
M. Armas Padilla,$^{1,2}$\thanks{E-mail: m.armaspadilla@iac.es}
G. Ponti,$^{3}$
B. De Marco,$^{4}$
T. Mu\~noz-Darias$^{1,2}$
and F. Haberl$^{3}$
\\
$^{1}$Instituto de Astrof\'isica de Canarias (IAC), V\'ia L\'actea s/n, La Laguna 38205, S/C de Tenerife, Spain\\
$^{2}$Departamento de Astrof\'isica, Universidad de La Laguna, La Laguna, E-38205, S/C de Tenerife, Spain\\
$^{3}$Max--Planck--Institut f\"ur extraterrestrische Physik, D-85748 Garching, Germany\\
$^{4}$Nicolaus Copernicus Astronomical Center, Polish Academy of Sciences, Bartycka 18, PL-00-716 Warsaw, Poland
}

\date{Accepted XXX. Received YYY; in original form ZZZ}

\pubyear{2017}

\begin{document}
\label{firstpage}
\pagerange{\pageref{firstpage}--\pageref{lastpage}}
\maketitle

\begin{abstract}
We report on a detailed study of the spectral and temporal properties of the neutron star low mass X-ray binary \SLX, which is located only $\sim$1$\degr$ away from Sgr A$^{*}$.  The system is expected to have a short orbital period, even within the ultra-compact regime, given its persistent nature at low X-ray luminosities and the long duration thermonuclear burst that it has displayed. We have analysed a  \suzaku\ (18 ks) observation and an \xmm\ (39 ks) observation taken 7 years apart. We infer (0.5--10~keV) X-ray luminosities in the range 3--6 $\times 10^{35}\lum$, in agreement with previous findings. The spectra are well described by a relatively cool ($kT_{\rm bb}$ = 0.5~keV) black body component plus a Comptonized emission component with $\Gamma \sim$1.5--1.7. These values are consistent with the source being in a faint hard state, as confirmed by the $\sim 20$ per cent fractional root mean square amplitude of the fast variability (0.1--7 Hz) inferred from the \xmm\ data. The electron temperature of the corona is $\gtrsim 7$ keV for the \suzaku\ observation, but it is measured to be as low as $\sim$2~keV in the \xmm\ data at higher flux. The latter is significantly lower than expected for systems in the hard state.
We searched for X-ray pulsations and imposed an upper limit to their semi-amplitude of 2 per cent (0.001 -- 7 Hz). Finally, we investigated the origin of the low frequency variability emission present in the \xmm\ data and ruled out an absorption dip origin. This constraint the orbital inclination of the system to $\lesssim$65$\degr$ unless the orbital period is longer than 11 hr (i.e. the length of the \xmm\ observation).

\end{abstract}

\begin{keywords}
accretion, accretion discs -- stars: individuals: (\SLX) -- stars: neutron -- X-rays: binaries
\end{keywords}



\section{Introduction}\label{sec:intr}

Low-mass X-ray binaries (LMXBs) are formed by a neutron star (NS) or a black hole (BH) accreting material from a low mass star via Roche lobe overflow. Transient systems spend long periods of time (years to decades) in a dim (\lx$<$10$^{33}\lum$) quiescent state, during which none or very little accretion  takes place. However, these dormant states are sporadically interrupted by short (weeks to months) accretion outburst events during which the X-ray luminosity increases by several orders of magnitude. Persistent sources, on the other hand, always show high X-ray luminosities and never go to quiescence \citep[see e.g.][]{VanderKlis2006,Belloni2011,Corral-Santana2016,Tetarenko2016}.

To understand the accretion flow properties at different accretion regimes has been a continuous challenge in the study of LMXBs. In particular, the luminosity regime below 1 per cent of the Eddington luminosity (\ledd) has not been intensively explored until a few years ago, due to the sensitivity limitation of the past X-ray missions. In addition, transient LMXBs typically spend very brief periods of times within the range \lx$\sim$10$^{34}$-10$^{36}\lum$, which reduces the window of opportunity for obtaining high-quality data.  On the other hand, systems  persistently accreting at such low luminosities are very rare, but they provide a unique opportunity to investigate these accretion regimes \citep{IntZand2002,DelSanto2007,Degenaar2010,ArmasPadilla2013b}. Just their mere persistent behaviour at such low accretion rates challenges the standard and well tested transient/persistent paradigm explained by the disc instability model (\mbox{\citealt{Lasota2001}},\mbox{\citealt{Coriat2012}}); a possible solution requiring  they having very short orbital periods, even within the ultra-compact regime \citep{Nelemans2010a}. 

\SLX\ is a neutron star (NS) low mass X-ray binary located in the Galactic center region ($\alpha$= 17$^{\rm h}$ 40$^{\rm m}$ 42\fs83  $\delta$= --28\degr 18\arcmin 08\farcs4; \citealt{Tomsick2007}) that was discovered with the \textit{Spacelab-2} observatory in 1985 \citep{Skinner1987}. The source has been subsequently detected by several monitoring programs with a flux of a few times 10$^{-11} \flux$ depending on the epoch and X-ray band. To date, four intermediate-long type I X-ray burst  have been reported from this source. The first one, which revealed the NS nature of the accretor, lasted $\sim$15 min  \citep{IntZand2002}, whilst the remaining three were 20--30 min long  and suggest a burst recurrence time of $\sim$~86~d \citep{Falanga2008}. At least one of these events shows photospheric radius expansion, from which, using the Eddington luminosity for a pure helium atmosphere, a distance of  7.3~kpc was derived \citep{Falanga2008}. This translates into a persistent luminosity of L$_\mathrm{X}$ (0.5--10 keV) $\sim$6--9$\times 10^{35}$ \ergs (i.e. $\sim$1.6--2.4 $\times 10^{-3}$\ledd; \citealt{IntZand2002}). 
The fact that \SLX\ has exhibited only intermediate long bursts, the 90~d recurrence time and the persistent nature of the source at such low luminosities makes it an ultra-compact X-ray binary (UCXB) candidate \citep{IntZand2007, Falanga2008}.

In this work we present X-ray spectroscopy from \suzaku\ and \xmm\ together with timing analysis from the latter facility, resulting in the most detailed soft X-ray analysis of the source up to date.

\section{Observations and data reduction}\label{sec:obs}

\subsection{\suzaku\ data}\label{subsec:suzaku}

The \suzaku\ observatory \citep{Mitsuda2007} observed \SLX\ for 18~ks on 2009 March 11. The X-ray Imaging Spectrometer (XIS; \citealt{Koyama2007}) was operated in the normal mode using the full window option (time resolution of 8~s).  We used the \textsc{heasoft} v.6.20 software and \suzaku\ Calibration Database (CALDB) to analyse our data. Following  \citet{ArmasPadilla2017}, to which we refer the reader for further details on the data reduction, we created light curves and spectra with the \textsc{ftool} \verb'xselect' using circular regions of 100~arcsec radius centred on the source and on a source-free part of the CDD for extracting source and background events, respectively. We used the \verb'xisrmfgen' and \verb'xisarfgen' commands to produce the response matrix and ancillary response files. We applied the \verb'aepileupcheckup.py' script \citep{Yamada2012} to verify that the pile-up fraction is less than 1 per cent at any radius of the point spread functions. We combined spectra and response files of the two Front-Illuminated (FI) detectors (XIS0 and XIS3) with the \verb'addascaspec' task in order to maximize the signal-to-noise ratio. Because of the discrepancy with the FI-XISs detectors, probably by cross calibration uncertainties, we excluded the  backside-illuminated camera XIS-1 spectrum from our analysis. In the Hard X-ray Detector (HXD; \citealt{Takahashi2007}) the source was only significantly detected by the PIN silicon diodes. We made use of the script \verb'hxdpinxbpi' to generate the source and the non-X-ray background spectra using the cosmic X-ray background correction option. We used the response files provided in the \suzaku\ CALDB.

\subsection{\xmm\ data}\label{subsec:xmm}

\SLX\ was observed on 2016 March 6 with the \xmm\ observatory \citep{Jansen2001} as part of the Galactic center lobe observations which extend the Galactic center X-ray scan \citep{Ponti2015}. During the 39~ks observation with \SLX\ in the field of view, both MOS \citep{Turner2001} and PN \citep{Struder2001} detectors of the European Photon Imaging Camera (EPIC) were operated in imaging (full-frame window) mode. In the MOS1 detector the source is placed between the CCD--1 and the dead CCD--6, therefore we did not include these data in our study. We used the Science Analysis Software (\textsc{sas}, v.16.0.0) to obtain calibrated events and scientific products.

We filtered episodes of flaring background by excluding data with count rates $>$~0.22~$\cnts$ at energies $>$10~keV and $>$~0.5~$\cnts$ at energies 10--20~keV for the MOS and PN cameras, respectively. Both detectors were affected by pile-up (net count rate of $\sim$3~$\cnts$ and $\sim$7~$\cnts$ when pile-up starts to be an issue when exceeding 0.5 and 2~$\cnts$, respectively). In order to mitigate the effect of pile-up in our data we used an annular extraction region centred at the source position with 60 arcsec outer radius and 10 arcsec inner radius. To extract MOS2 background events we used an annulus centred at the source with 100 and 300 arcsec of inner and outer radius, respectively. In the case of PN background events, we used a circular region of 130 arcsec radius placed in the central part of the detector in order to avoid the EPIC internal 'quiescence' background issues (see the \xmm\ users' handbook\footnote{\url{https://xmm-tools.cosmos.esa.int/external/xmm_user_support/documentation/uhb/XMM_UHB.html}}). Light curves and spectra, as well as associated response matrix files (RMFs) and ancillary response files (ARFs) were generated following the standard analysis threads\footnote{\url{https://www.cosmos.esa.int/web/xmm-newton/sas-threads}}. We rebinned the spectrum in order to include a minimum of 25 counts in every spectral channel and avoiding to oversample the full width at half-maximum of the energy resolution by a factor larger than 3.

\section{Analysis and results}\label{sec:anares}

\subsection{Temporal analysis}\label{subsec:temp}

We took advantage of the length and good enough time resolution of the  \xmm\ data to carry out a detailed timing analysis.  The \suzaku\ XIS data were not included in this analysis as their time resolution is too low.

\begin{figure}
\begin{center}
\includegraphics[keepaspectratio,width=\columnwidth, trim=0.0cm 0.0cm 0.0cm 0.0cm]{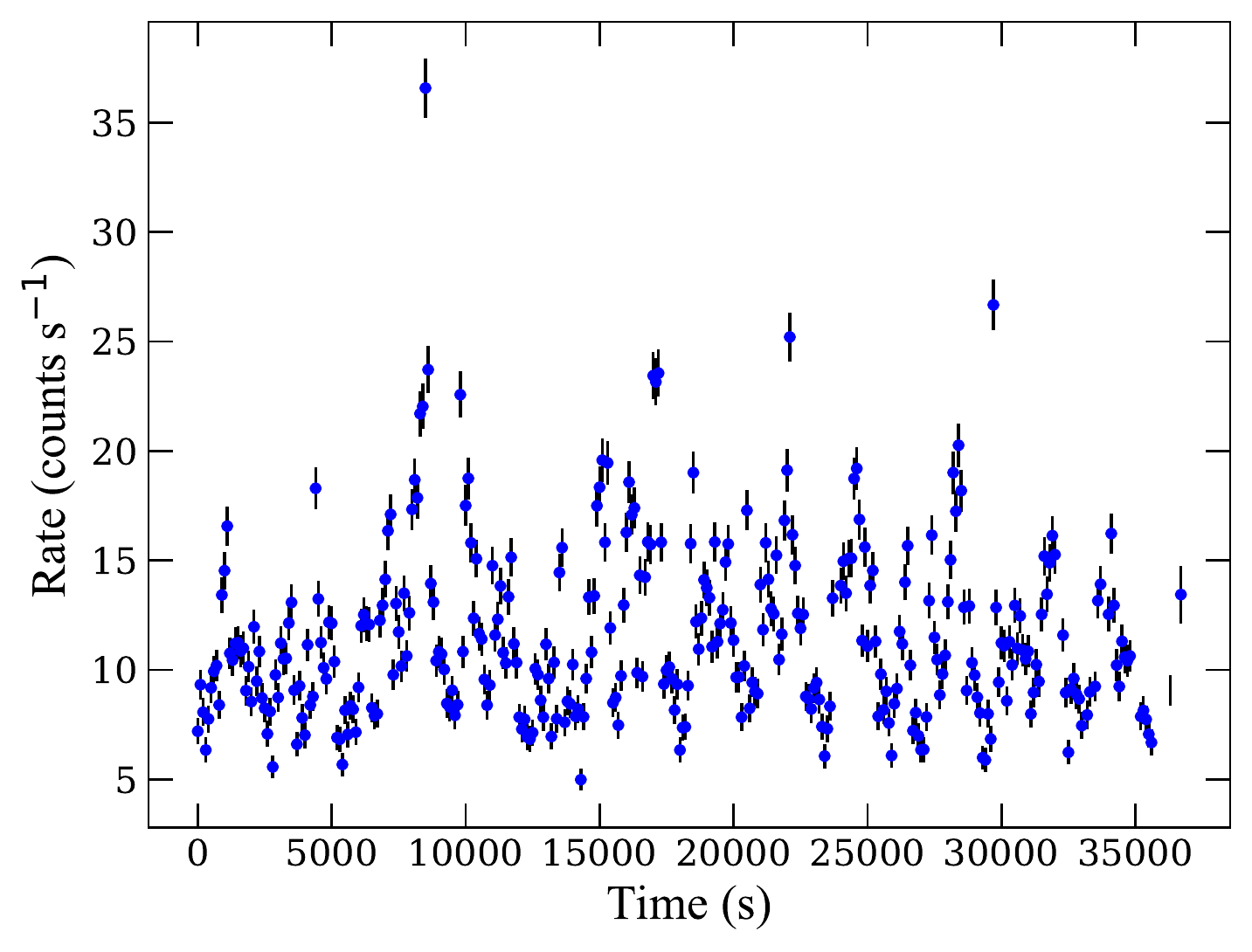}\\
\caption{EPIC-pn 1-10~keV light curve at 100~s resolution.}
\label{fig:LC}
\end{center}
\end{figure}

\begin{figure}
\begin{center}
\includegraphics[keepaspectratio,width=\columnwidth, trim=0.0cm 0.0cm 0.0cm 0.0cm]{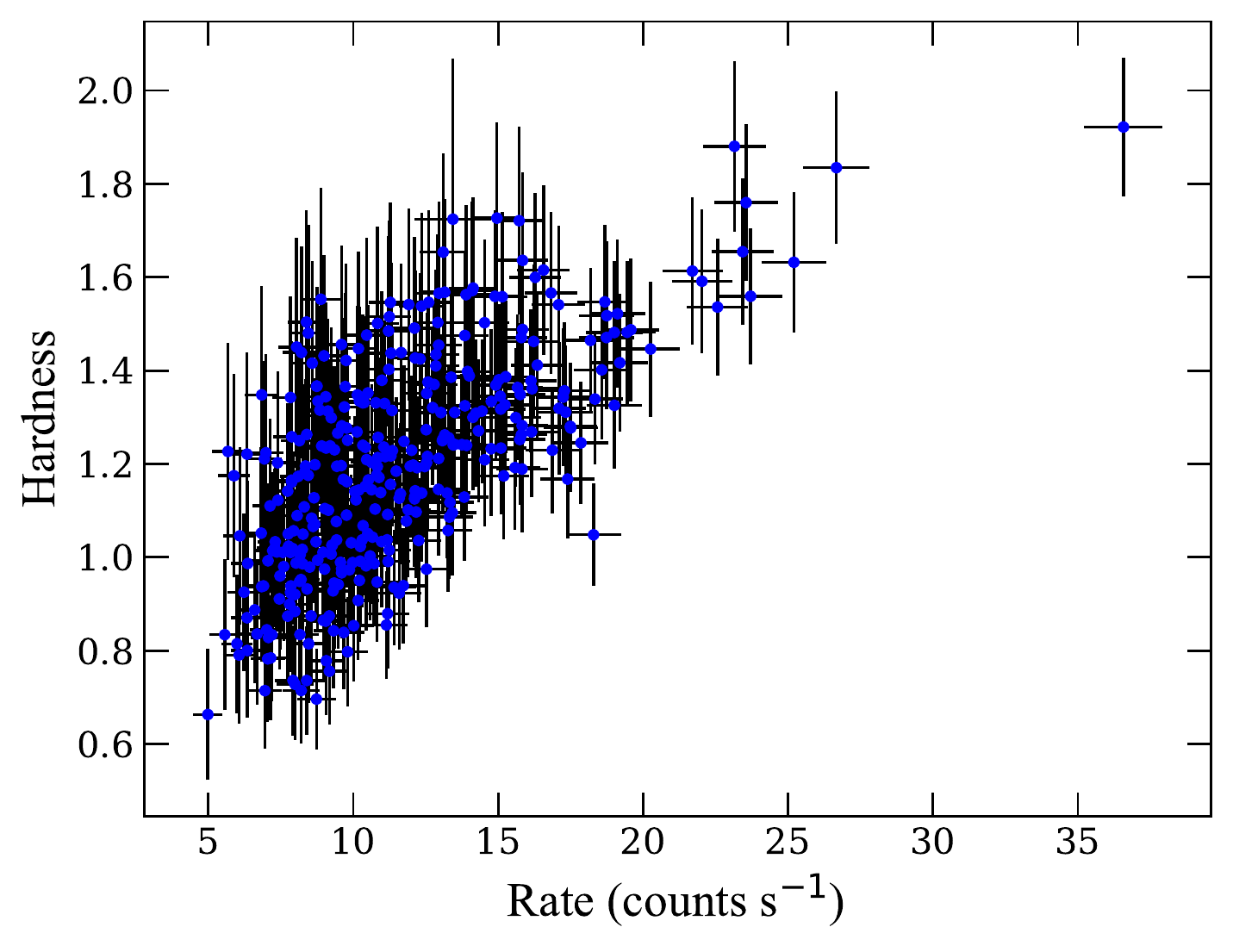}\\
\caption{Hardness ratio (2.5-10~keV / 1-2.5~keV) versus intensity diagram.}
\label{fig:HR}
\end{center}
\end{figure}

In a first step we investigated the 1-10~keV EPIC-pn light curve using a 100s bin size, which reveals the presence of significant variability (Fig. \ref{fig:LC}). The origin of this variability  can be (broadly speaking) either accretion or absorption related. The latter is a very intriguing possibility as  absorption dips allow measuring the orbital period of relatively high-inclination systems (with $i\gtrsim65^{\circ}$, \citealt{White1985}; see also \citealt{Cantrell2010} for an accurate inclination measurement in a non--dipping source). Indeed, dips are produced by the obscuration of the central X-ray source by orbital phase locked disc material, most likely in the disc bulge, which is the result of the gas stream from the companion impacting on the outer accretion disc rim. As the effect of obscuration by lowly ionised material is more conspicuous in the soft energy band, spectral hardening should be observed in correspondence with dip events \citep[e.g.][]{White1985, Frank1987, Kuulkers2013, Ponti2016,DeMarco2015}. To test this possibility, we plot in Fig. \ref{fig:HR} the corresponding hardness ratios, computed as the count-rate ratio of the (hard) 2.5--10~keV band to the (soft) 1.0--2.5~keV band, versus intensity. We observe that the hardness ratio is correlated with the X-ray flux, which rules out obscuration dips as the origin of the observed low frequency variability. Given the length of the \xmm\ observations (39 ks) we conclude that \SLX\ has an orbital inclination $\lesssim$65$\degr$ unless the orbital period of the source is longer than $\sim11$~h, which is most likely infeasible given the X-ray behaviour of the source.

In a second step, we searched for periodic modulations by creating Lomb-Scargle periodograms \citep{Scargle1982} using the 0.3$-$8.0 keV EPIC-pn light curve. We investigated periods, linearly spaced in frequency, in the range $3 \times 0.073$~s (i.e. the CCD readout frame time) to 1000~s, using frequency resolution of $\Delta f = 1/T = 2.7\times10^{-5}$~Hz with $T$ the duration of the observation. We did not detect any significant signal and derived upper limits for sinusoidal modulations at several of the highest peaks in the Lomb-Scargle power. Given the high count rate and the long observation we are able to derive a stringent upper limit of 2 per cent for the semi-amplitude of possible sinusoidal modulations such as pulses from the NS.

\begin{figure}
\begin{center}
\includegraphics[keepaspectratio,width=\columnwidth, trim=0.5cm 0.0cm 0.0cm 0.0cm]{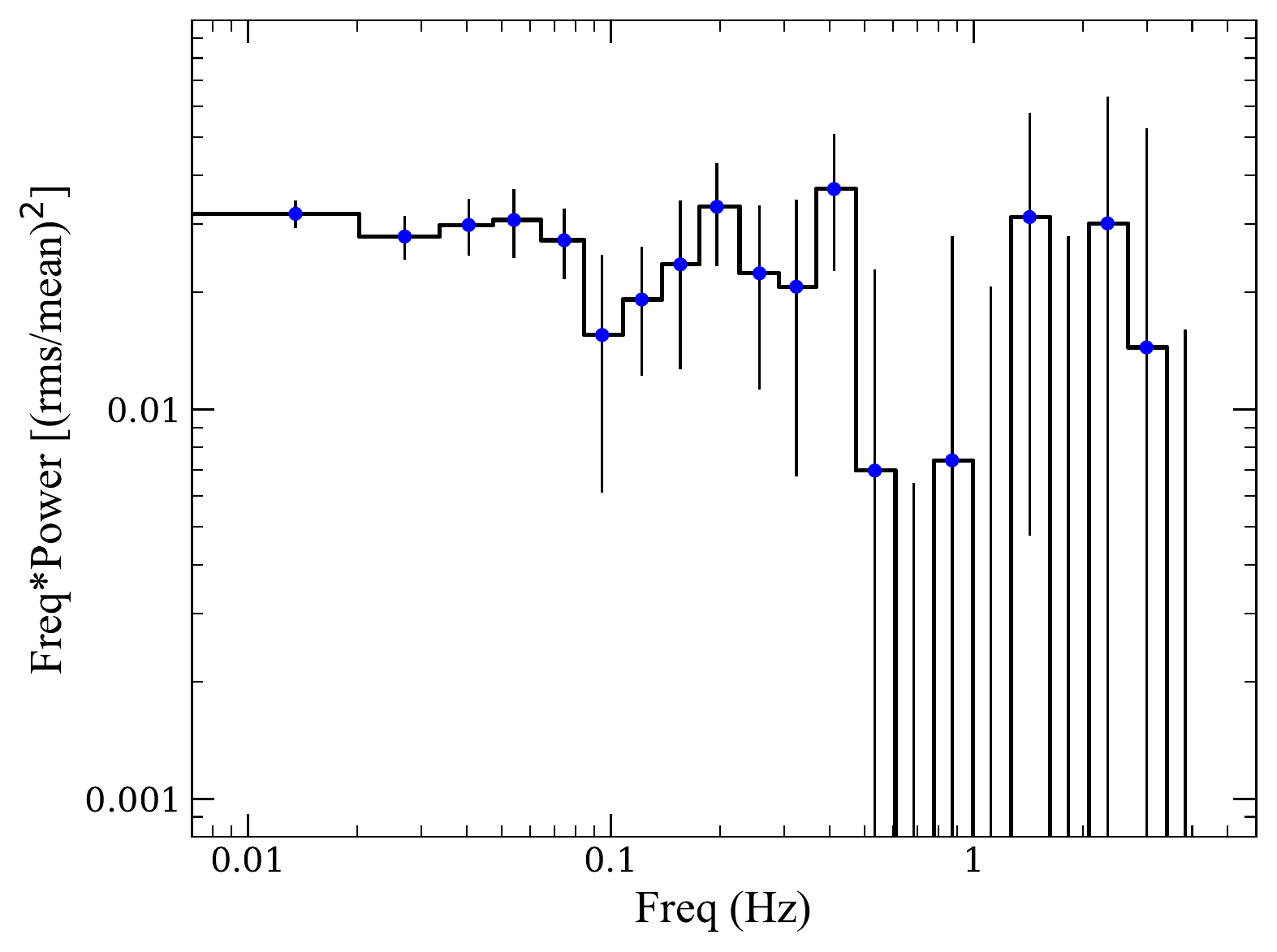}\\
\caption{The PSD of SLX 1737-282 in the 2-10 keV energy band. The PSD is Poisson noise-subtracted and displayed using the fractional root-mean-square normalization \citep{Miyamoto1991}.}
\label{fig:pow_spec}
\end{center}
\end{figure}

Finally, and following \citet{Munoz-Darias2011a,Munoz-Darias2014},  we estimated the fractional root mean square (rms) of the fast variability amplitude \citep[e.g.][]{Vaughan2003,Ponti2012} to infer the accretion state of the source. To this end, we extracted the EPIC-pn light curve in the energy band 2-10~keV using 74~ms time bins. We divided the light curve into 74~s segments and computed the power spectral density function (PSD) for each of them. The contribution of Poisson noise to the total variability was estimated  by fitting a constant to the high-frequency ($\nu\geq2$~Hz) part of each PSD. The derived Poisson noise contribution was subsequently subtracted for every frequency. Each PSD was normalized adopting the squared fractional rms normalization (Miyamoto et al. 1991). The intrinsic fractional rms was estimated by integrating the Poisson noise-subtracted, normalized PSDs in the frequency range 0.1-7~Hz, and averaging the resulting values. The average PSD is shown in Fig. \ref{fig:pow_spec}. We infer a fractional rms of $22\pm3$ per cent (where the error corresponds to the $1\sigma$ uncertainty). Comparing this value with those typically displayed by BHs (e.g. \citealt{Munoz-Darias2011a,Munoz-Darias2011b}, \citealt{Heil2012}) and NSs \citep{Munoz-Darias2014} LMXBs during active phases, the inferred value is strongly indicative of \SLX\ being in the hard state during the \xmm\ observation. 

\subsection{Spectral analysis}\label{subsec:sepc}

We used \textsc{xspec} (v.12.9.1; \citealt{Arnaud1996}) to analyse both the \suzaku\ and \xmm\ spectra. In order to account for interstellar absorption, we used the Tuebingen-Boulder Interstellar Medium absorption model (TBABS in \textsc{xspec}) with cross-sections of \citet{Verner1996} and abundances of \citet{Wilms2000}. In this work we assume a distance of 7.3~kpc \citep{Falanga2008}, NS mass and radius of 1.4~M\sun\  and 10~km, respectively, and an orbital inclination of 60\degr.

\subsubsection{\suzaku\ spectra}
We simultaneously fit the 1-10~keV FI-XISs and 15-30~keV HXD/PIN spectra by using tied spectral parameters. HXD/PIN data above 30 keV were excluded since they do not exceed the background level (at 3$\sigma$). In order to avoid the silicon K-edge and gold-M-edge calibration uncertainties, we exclude the 1.7-2.4~keV energy range. We included a constant factor (CONSTANT) fixed to 1 for XIS spectra and 1.16 for HXD spectra to account for cross-calibration uncertainties between the instruments\footnote{Suzaku Memo 2008-06 at \url{http://www.astro.isas.jaxa.jp/suzaku/doc/suzakumemo}}.

In a  first attempt to fit the spectra we used a simple 2-component model from the \textit{hybrid} modelling proposed by \citet{Lin2007} to fit NS spectra. We combined a soft component to account for either emission from a relatively cold accretion disc or from the NS surface/boundary layer, and a hard component to model the inverse-Comptonized emission from the Corona. Thus, we used a multicolour disc (MCD) or single black body (BB) plus a thermally Comptonized continuum model (i.e. DISKBB+NTHCOMP and BBODYRAD+NTHCOMP in \textsc{xspec}; \citealt{Mitsuda1984,Makishima1986, Zdziarski1996, Zycki1999}). 

\begin{figure}
\begin{center}
\includegraphics[keepaspectratio,width=\columnwidth, trim=0.5cm 0.5cm 2.0cm 1.0cm]{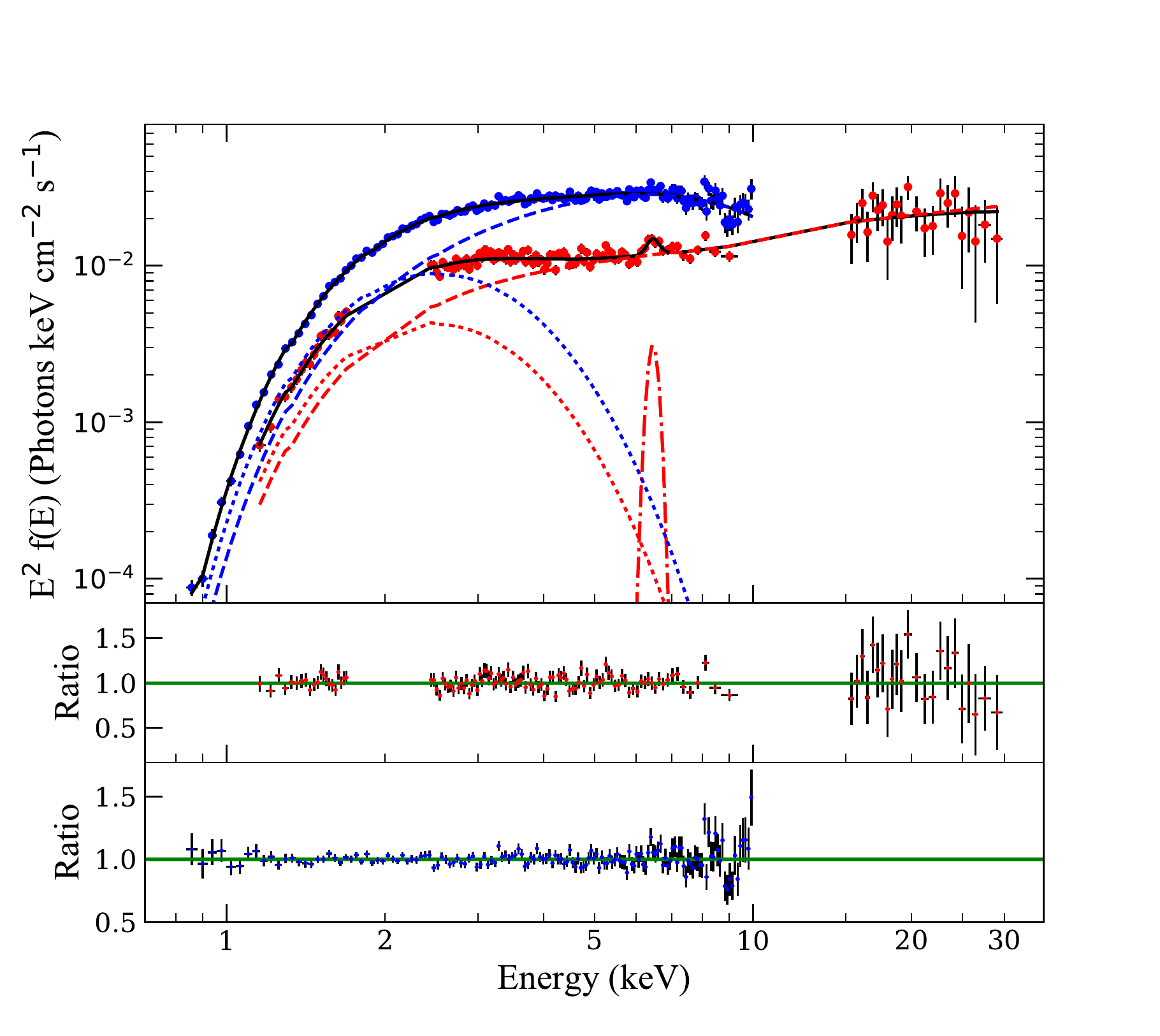}\\
\caption{ Unfolded \suzaku\ (red) and \xmm\ (blue) spectra and data-to-model ratio using BBODYRAD+NTHCOMP model. The full fit is shown as a solid line, the black body component as a dotted line, the Comptonized component as a dashed line and the \suzaku\ Gaussian component as a dot-dashed line.}
\label{fig:spectra}
\end{center}
\end{figure}

Although both models return acceptable fits [\qhis$\cong$153 for 129 degrees of freedom (dof)], the MCD+Comptonization model yields results that are not physically acceptable. Indeed, the corrected inner disc radius inferred from the DISKBB normalization component (N$_{\rm diskbb}\sim$~5) is smaller than the size of the NS ($R_{\rm disk}<$10~km, applying correction of \citealt{Kubota1998}, see also \citealt{ArmasPadilla2017} and references therein). In addition, the temperature obtained is  $kT_{\rm diskbb}$=0.67$\pm$0.08~keV, which is a factor $\sim$2-3 higher than the disc temperatures observed in other NSs \citep[e.g.][]{ArmasPadilla2017} and black holes \citep[e.g.][]{Miller2006b,Munoz-Darias2013,ArmasPadilla2014a,Shidatsu2014,Plant2015} in the hard state at similar luminosity ranges. Therefore, we select the latter model (BBODYRAD+NTHCOMP) to fit our data. This model provides an overall good fit (\qhis=153.3 for 129 dof) but it leaves residuals around $\sim6.4$ keV consistent with Fe~K$_{\alpha}$ emission. Thus, we added a Gaussian (GAUSS) component to the modelling, which improved significantly the fit (\qhis=133.12 for 126 dof, F-test probability $\sim$10$^{-4}$). The line has a central energy of $E_{\rm l}$=6.4$\pm$0.1~keV,  $\sigma_{\rm l}$=0.16$\pm$0.1~keV, therefore barely resolved ($\Delta$\qhis$\sim$5.78 if $\sigma_{\rm l}$ is fixed to 0), and an equivalent width of 110$\pm$50~eV. The Fe~K$_{\alpha}$ line is typically the most prominent feature of a reflection component from optically thick material \citep[e.g.][]{Cackett2009,Ng2010}. We checked for the presence of a neutral reflection continuum associated with the Fe~K$_{\alpha}$ line by replacing the Gaussian with the neutral Compton reflection model (PEXNOM in \textsc{xspec}; \citealt{Nandra2007}). This provides no significant changes in the fit ($\Delta$\qhis=2.8 for 2 dof) indicating that this line could be produced by reflection from either optically thick or thin material \citep{Garcia2015,Nayakshin2000}. The best fit results are reported in Table \ref{tab:res} and Fig. \ref{fig:spectra} with uncertainties given at 90 per cent confidence level.

We obtain an equivalent hydrogen column density (\Nh) for the absorbing material of (1.9$\pm$0.2)$\times$10$^{22}$~\nh, consistent with the value reported by \citet{IntZand2002}. The temperature of the black-body component ($kT_{\rm bb}$=0.49$\pm$0.04~keV) is within the typical range of temperatures measured for other NS LMXBs in the hard state \citep[e.g.][]{Campana2014}. We infer an emission radius ($R_{\rm bb}$) of $\sim$3~km from the blackbody normalization assuming a spherical geometry, which might indicate that the emission arises from a small region of the NS (e.g. from an equatorial belt in the orbital plane or from the magnetic poles;  \citealt{Barret2001,Lin2007,Matsuoka2013}). Nevertheless, we note that it is not straightforward to obtain physically meaningful $R_{\rm bb}$ values from this kind of modelling. Also, we did not apply any correction for the photons scattered by the Corona which might increase the obtained values by a factor<2 \citep[see][]{Kubota2004}. The Comptonization asymptotic power-law photon index ($\Gamma$) is 1.69$\pm$0.09. The temperature of the electron corona is not constrained, and has a lower-limit of  $kT_{\rm e} >$~7.4~keV. The inferred  0.5--10~keV unabsorbed flux is ($4.6\pm0.1)\times10^{-11}~\flux$, which corresponds to \lx=($2.93\pm 0.06)\times10^{35}~\lum$. The thermal component accounts for 30 per cent of the total flux in this band.  Similarly, in the 0.5--30 keV band we measure (7.7$\pm0.2$)$\times10^{-11} \flux$ [\lx(0.5--30 keV)=(4.9$\pm$0.1)$\times$10$^{35}~\lum$] and a thermal contribution of 18 per cent.

\subsubsection{\xmm\ spectrum}

For clarity purposes, we report the results of the fit of EPIC-pn (the instrument with the larger effective area) data only. Nevertheless, we note that results obtained from MOS2 as well as the combination of PN and MOS spectra yield consistent results.

We fit the 0.8--10~keV spectrum with the same model used for the \suzaku\ data, but excluding the Gaussian component  since it was not required by the fit. Therefore, we used BBODYRAD+NTHCOMP, which produces an acceptable fit (\qhis=148.04 for 148 dof). We infer a 0.5--10~keV unabsorbed flux of (10.3$\pm$0.3)$\times$10$^{-11}\flux$ which translates into  L$_\mathrm{X}$ = (6.6$\pm$0.2)$\times$10$^{35}\lum$, that is, a factor of two more luminous than the \suzaku\ spectra.
The black body parameters, $kT_{\rm bb}$=0.53$\pm$0.05~keV and $R_{\rm bb}$=4.1$\pm$0.3~km are consistent with those obtained with \suzaku\ while $\Gamma$ is slightly harder (1.49$\pm$0.4). In this case the electron temperature was constrained to be atypically small ($kT_{\rm e}$=2.0$\pm$0.4~keV; see section \ref{sec:Disc}).  Finally, and in agreement with the \suzaku\ spectrum we obtain \Nh=(1.9$\pm$0.1)$\times$10$^{22}$~\nh (see Tab. \ref{tab:res}).

\subsubsection{Exploring alternative spectral scenarios}
Recent studies have shown that the 3-component model from the \textit{hybrid} model \citep{Lin2007} adequately fits both hard and soft state NS spectra when high-quality data (including broad-band coverage and soft energies) are available (\citealt{ArmasPadilla2017}, Ponti et al. 2017, submitted). Thus, we explored the 3-component model (DISKBB+BBODYRAD+NTHCOMP) for fitting our spectra. This modelling yields good fits (\qhis=129.02 for 124 dof and \qhis=147.19 for 146 dof for \suzaku\ and \xmm\, respectively) with spectral values within the usual ranges for NSs in the hard state. However, the fits become degenerated, as the extra thermal component is not strictly required by the data (F-test probability $\sim$0.1). For reference, the disc temperature upper limits are $kT_{\rm diskbb}<$0.5~keV and $<0.7$~keV for \suzaku\ and \xmm, respectively. In the same way, we also examined the most simplistic model, a single Comptonization model (NTHCOMP). This model does not provide a good fit for the \suzaku\ spectra (\qhis=226.62 for 130 dof).  On the other hand,  although it returns an acceptable fit for the \xmm\ data (\qhis=161.54 for 149 dof), an F-test indicates that the fit would significantly improve by adding the thermal component (F-test probability $\sim$10$^{-4}$).  Additionally, the inferred $\Gamma\sim$2 is softer than typically observed in the hard state (but see \citealt{Wijnands2015} for only power-law fits).

\begin{table}
\centering
\caption{Fitting results for the BBODYRAD+NTHCOMP model for the \suzaku\ and \xmm\ observations. Uncertainties are expressed at 90 per cent confidence level.}
\begin{threeparttable}
\begin{tabular}{ l c c }
\hline
  Component 					&\suzaku 			& \xmm	 \\		 			
\hline
\Nh\ ($\times 10^{22}$ \nh)	& 	1.9$\pm$0.2 		&  1.9$\pm$0.1			 		\\
$kT_{\rm bb}$ (keV)			& 	0.51$\pm$0.04 	&  0.53$\pm$0.05					\\
$N_{\rm bb}$	/$R_{\rm bb}$ (km)& 	18$^{+8}_{-5}$/3.1$^{+0.6}_{-0.4}$	&  32$^{+6}_{-7}$/4.1$\pm$0.3 		 \\
$\Gamma$				 		& 	1.68$\pm$0.08   & 1.49$^{+0.3}_{-0.4}$		   \\
$kT_{\rm e}$ (keV)			& 	$>$7.4	 		& 2.0$^{+0.4}_{-0.3}$	 	\\
$N_{\rm nthcomp}$ ($\times 10^{-3}$) & 	2.17$\pm$0.5		& 3.7$\pm$3  	 			  	\\
$E_{\rm gaus}$ (keV)			&	6.4$\pm$0.1		&	--						\\
$\sigma_{\rm gaus}$ (keV)		& 0.16$^{+0.1}_{-0.08}$	&	--						\\
$k_{\rm gaus}$ ($\times 10^{-5}$ photons cm$^{-2}$ s$^{-1}$) 	& 3$\pm$1		&	--						\\
\qhis\ /dof 				& 	133.12/126 		& 148.04/148 					  \\
 &&\\
\multicolumn{1}{r}{(0.5--10 keV)} &&\\
\Fx\ ($\times10^{-11} \flux$)				&	4.6$\pm$0.1		& 9.77$\pm$0.2			 	\\
\lx$^{\rm a}$ ($\times10^{35}\lum$)		&	2.93	$\pm$0.06	& 6.2$\pm$0.1				 \\
BB fraction					&	30				&	27						 \\
NTHCOMP fraction				&	70				&	73						\\
 &&\\
\multicolumn{1}{r}{(0.5--30 keV)} &&\\
\Fx\ ($\times10^{-11} \flux$)				&	7.68$\pm$0.2		& --							\\
\lx$^{\rm a}$ ($\times10^{35}\lum$)		&	4.89$\pm$0.1		& --							\\
BB fraction					&	18				&	--						\\
NTHCOMP fraction				&	82				&	--						\\

\hline
\end{tabular}
\begin{tablenotes}
\item[a]{Unabsorbed luminosity assuming a distance of 7.3~kpc.}

\end{tablenotes}
\label{tab:res}
\end{threeparttable}
\end{table}

\section{Discussion}\label{sec:Disc}

\SLX\ is one of the few known LMXB persistently accreting at luminosities lower than \lx$\sim$0.01\ledd\ \citep[e.g.][]{IntZand2005,DelSanto2007,Chelovekov2007,Degenaar2012d,ArmasPadilla2013b,Degenaar2017}. In order to study the spectral and timing properties in this regime, high-quality observations resulting from long exposures, such as those presented here, are essential.
In this work we report on two observations obtained with \suzaku\ and \xmm\  over a time span of 7 years. The 0.5--10~keV X-ray luminosity during these observations was in the range 3--6$\times$10$^{35}\lum$, which is consistent with previous measurements \citep{IntZand2002, Falanga2008}. This implies that the NS is still continuously accreting  at $10^{-3}$ \ledd. \SLX , which is located $\sim$1.3\degr\ above the Galactic center, is absorbed by a column density of neutral material of \Nh=1.9($\pm$0.1)$\times$10$^{22}$~\nh. This is a factor 8--9 times lower compared to the value observed for Sgr~A$^{*}$ and its immediate surroundings \citep{Ponti2017}. 

Our timing analysis does not reveal the presence of dipping phenomena, which would eventually allow us to constrain the orbital period of the system and test the suspected ultra-compact nature of \SLX. However, we are able to discard flux pulsations with semi-amplitudes larger than 2 per cent the observed flux. Moreover, from the \xmm\ observation we are able to measure an integrated fractional rms (0.1--7 Hz) of $\sim20$ per cent, which indicates that the source was in the hard state \citep{Munoz-Darias2014}  during this epoch, and strongly suggests a similar scenario for the \suzaku\ data given its lower associated luminosity. 

The spectral analysis of the source is consistent with the above picture, as the X-ray spectrum is well modelled by a combination of a thermal component with $kT_{\rm bb}\sim$0.5~keV, that contributes $\sim$30 per cent to the 0.5--10~keV flux, and a thermally Comptonized component with $\Gamma\sim$1.5--1.7 . \xmm\ observations of several NS LMXB at low-luminosities -- both transient \citep[e.g.][]{ArmasPadilla2011,Degenaar2013,Chakrabarty2014,Arnason2015} and persistent \citep[e.g.][]{ArmasPadilla2013b,Degenaar2017} systems -- have been successfully modelled with this 2-component model. However, the black body temperature inferred here is slightly higher than typically observed in other persistent, low-luminosity systems with $kT_{\rm bb}\sim$0.3~keV \citep{ArmasPadilla2013b}. Still, these objects have luminosities which are up to one order of magnitude lower. Indeed, the $kT_{\rm bb}$ temperature obtained for IGR~J17062-6143, quasi-persistent at  $\sim$1.6$\times$10$^{35}\lum$, is similar to that measured here for \SLX\ \citep{Degenaar2017}. The radius obtained from the black body normalization (Rbb$\sim$3--4~km) is smaller than the NS radius, and this might indicate that the emission arises from a reduced region of the NS surface. This could be explained by a boundary layer with a belt shape, in which the gas from a geometrically thin disc is accreted towards the NS equatorial region. However, this accretion geometry seems unlikely in the mid-to-low luminosity phase of the hard state. On the other hand, accretion through the magnetic poles could also account for the small black body normalization, but we do not detect coherent X-ray pulsations \citep[but see ][for a possible non detection justification]{Matsuoka2013}. In any case, we note that the radius obtained by this modelling is not physically meaningful and it could be underestimated by a factor of up to 2 (see \ref{subsec:suzaku}).

While the overall spectral parameters are consistent between the \xmm\ and the \suzaku\ observations, the temperatures measured for the Comptonizing electron cloud are found to be different. The (broader band) \suzaku\ spectrum obtained at lower luminosity yields only a lower-limit of $kT_{\rm e}$>7.4~keV, whilst the fit of the \xmm\ data requires a very low electron temperature ($kT_{\rm e}\sim$2~keV). 
The temperature of the corona in NS systems is typically observed to change through the outburst according to the different accretion states. In the soft state, the Corona is characterised by a cold temperature ($kT_{\rm e}\lesssim$5~keV) and an optical depth $\gtrsim$5, whereas in the hard state $kT_{\rm e}\gtrsim$10~keV is typically observed as well as lower optical depths of $\sim2$ (\citealt{Lin2007}, \citealt{ArmasPadilla2017}; see also \citealt{Motta2009} for BHs). In addition, recent work by \citet{Burke2017} illustrates the differing Corona properties between BH and NS in the hard state. In particular, the electron temperature in NS systems is observed to be lower due to the additional cooling from seed photons coming from the boundary layer or the NS star surface. In any case, the inferred value from our \xmm\ observation is lower than typically observed for NS in the hard state. However, we note that the low luminosity regime has been scarcely explored and the broader spectral coverage of our \suzaku\ observation provides a result which is consistent with standard values.

\section{conclusions}
We have presented the most detailed soft X-ray study up to date of the persistent, low-luminosity neutron star X-ray binary \SLX.  Both the spectral and timing analysis of the source are consistent with \SLX\ being in a very faint hard state. The spectral parameters are in agreement with those derived from the handful of high-quality observations obtained for objects of this class. On the other hand, some observables, such as those related to the physical conditions of the Comptonizing corona remain unclear, as well as the ultra-compact nature of the source. Future multi-wavelength observations of \SLX\ and other persistent, low-luminosity LMXBs are required in order to get new insights on the accretion physics at work in this sub-luminous regime.

\section*{Acknowledgements}
We thank Chichuan Jin for discussion and help with modelling of the distortions induced by dust scattering. The research leading to these results has received funding from the European Union's Horizon 2020 Programme under the AHEAD project (grant agreement n. 654215). MAP and TMD acknowledge the hospitality of the Max-Planck-Institut f\"ur extraterrestriche Physik, where this work was carried out. MAP's research is funded under the Juan de la Cierva Fellowship Programme of the Ministry of Science and Innovation (MINECO) of Spain. BDM acknowledges support from the European Union's Horizon 2020 research and innovation programme under the Marie Sk{\l}odowska-Curie grant agreement No.665778 via the Polish National Science Center grant Polonez UMO-2016/21/P/ST9/04025. TMD acknowledges support via a Ramon y Cajal Fellowship (RYC-2015-18148). This research has made use of data obtained from the \suzaku\ satellite, a collaborative mission between the space agencies of Japan (JAXA) and the USA (NASA). The \xmm\ Galactic Lobe project is supported by the Bundesministerium f\"ur Wirtschaft und Technologie/Deutsches Zentrum f\"ur Luft- und Raumfahrt (BMWI/DLR, FKZ 50 OR 1408) and the Max Planck Society. \xmm\ is an ESA science mission with instruments and contributions directly funded by ESA Member States and NASA.



\bibliographystyle{mnras}
\bibliography{SLX1737_revision1.bbl} 

\begin{thebibliography}{}
\makeatletter
\relax
\def\mn@urlcharsother{\let\do\@makeother \do\$\do\&\do\#\do\^\do\_\do\%\do\~}
\def\mn@doi{\begingroup\mn@urlcharsother \@ifnextchar [ {\mn@doi@}
  {\mn@doi@[]}}
\def\mn@doi@[#1]#2{\def\@tempa{#1}\ifx\@tempa\@empty \href
  {http://dx.doi.org/#2} {doi:#2}\else \href {http://dx.doi.org/#2} {#1}\fi
  \endgroup}
\def\mn@eprint#1#2{\mn@eprint@#1:#2::\@nil}
\def\mn@eprint@arXiv#1{\href {http://arxiv.org/abs/#1} {{\tt arXiv:#1}}}
\def\mn@eprint@dblp#1{\href {http://dblp.uni-trier.de/rec/bibtex/#1.xml}
  {dblp:#1}}
\def\mn@eprint@#1:#2:#3:#4\@nil{\def\@tempa {#1}\def\@tempb {#2}\def\@tempc
  {#3}\ifx \@tempc \@empty \let \@tempc \@tempb \let \@tempb \@tempa \fi \ifx
  \@tempb \@empty \def\@tempb {arXiv}\fi \@ifundefined
  {mn@eprint@\@tempb}{\@tempb:\@tempc}{\expandafter \expandafter \csname
  mn@eprint@\@tempb\endcsname \expandafter{\@tempc}}}

\bibitem[\protect\citeauthoryear{{Armas Padilla}, Degenaar, Patruno, Russell,
  Linares, Maccarone, Homan  \& Wijnands}{{Armas Padilla}
  et~al.}{2011}]{ArmasPadilla2011}
{Armas Padilla} M.,  Degenaar N.,  Patruno a.,  Russell D.~M.,  Linares M.,
  Maccarone T.~J.,  Homan J.,   Wijnands R.,  2011, \mn@doi [MNRAS]
  {10.1111/j.1365-2966.2011.19308.x}, 417, 659

\bibitem[\protect\citeauthoryear{{Armas Padilla}, Degenaar  \& Wijnands}{{Armas
  Padilla} et~al.}{2013}]{ArmasPadilla2013b}
{Armas Padilla} M.,  Degenaar N.,   Wijnands R.,  2013, \mn@doi [MNRAS]
  {10.1093/mnras/stt1114}, 434, 1586

\bibitem[\protect\citeauthoryear{{Armas Padilla}, Wijnands, Altamirano, Mendez,
  Miller  \& Degenaar}{{Armas Padilla} et~al.}{2014}]{ArmasPadilla2014a}
{Armas Padilla} M.,  Wijnands R.,  Altamirano D.,  Mendez M.,  Miller J.~M.,
  Degenaar N.,  2014, \mn@doi [MNRAS] {10.1093/mnras/stu243}, 439, 3908

\bibitem[\protect\citeauthoryear{{Armas Padilla}, Ueda, Hori, Shidatsu  \&
  Mu{\~{n}}oz-Darias}{{Armas Padilla} et~al.}{2017}]{ArmasPadilla2017}
{Armas Padilla} M.,  Ueda Y.,  Hori T.,  Shidatsu M.,   Mu{\~{n}}oz-Darias T.,
  2017, \mn@doi [MNRAS] {10.1093/mnras/stx020}, 309, stx020

\bibitem[\protect\citeauthoryear{Arnason, Sivakoff, Heinke, Cohn  \&
  Lugger}{Arnason et~al.}{2015}]{Arnason2015}
Arnason R.~M.,  Sivakoff G.~R.,  Heinke C.~O.,  Cohn H.~N.,   Lugger P.~M.,
  2015, \mn@doi [ApJ] {10.1088/0004-637X/807/1/52}, 807, 52

\bibitem[\protect\citeauthoryear{Arnaud}{Arnaud}{1996}]{Arnaud1996}
Arnaud K.,  1996, in Jacoby G.,  Barnes J.,  eds,  Astronomical Society of the
  Pacific Conference Series Vol. 101, Astronomical Data Analysis Software and
  Systems V. p.~17

\bibitem[\protect\citeauthoryear{Barret}{Barret}{2001}]{Barret2001}
Barret D.,  2001, \mn@doi [Adv. Sp. Res.] {10.1016/S0273-1177(01)00414-8}, 28,
  307

\bibitem[\protect\citeauthoryear{Belloni, Motta  \& Mu{\~{n}}oz-Darias}{Belloni
  et~al.}{2011}]{Belloni2011}
Belloni T.~M.,  Motta S.~E.,   Mu{\~{n}}oz-Darias T.,  2011, Bulletin of the
  Astronmical Society of India, 39, 409

\bibitem[\protect\citeauthoryear{Burke, Gilfanov  \& Sunyaev}{Burke
  et~al.}{2017}]{Burke2017}
Burke M.~J.,  Gilfanov M.,   Sunyaev R.,  2017, \mn@doi [MNRAS]
  {10.1093/mnras/stw2514}, 466, 194

\bibitem[\protect\citeauthoryear{Cackett et~al.,}{Cackett
  et~al.}{2009}]{Cackett2009}
Cackett E.~M.,  et~al., 2009, \mn@doi [ApJ] {10.1088/0004-637X/690/2/1847},
  690, 1847

\bibitem[\protect\citeauthoryear{Campana, Brivio, Degenaar, Mereghetti,
  Wijnands, D'Avanzo, Israel  \& Stella}{Campana et~al.}{2014}]{Campana2014}
Campana S.,  Brivio F.,  Degenaar N.,  Mereghetti S.,  Wijnands R.,  D'Avanzo
  P.,  Israel G.~L.,   Stella L.,  2014, \mn@doi [MNRAS]
  {10.1093/mnras/stu709}, 441, 1984

\bibitem[\protect\citeauthoryear{Cantrell et~al.,}{Cantrell
  et~al.}{2010}]{Cantrell2010}
Cantrell A.~G.,  et~al., 2010, \mn@doi [ApJ] {10.1088/0004-637X/710/2/1127},
  710, 1127

\bibitem[\protect\citeauthoryear{Chakrabarty et~al.,}{Chakrabarty
  et~al.}{2014}]{Chakrabarty2014}
Chakrabarty D.,  et~al., 2014, \mn@doi [ApJ] {10.1088/0004-637X/797/2/92}, 797,
  92

\bibitem[\protect\citeauthoryear{Chelovekov \& Grebenev}{Chelovekov \&
  Grebenev}{2007}]{Chelovekov2007}
Chelovekov I.~V.,  Grebenev S.~A.,  2007, \mn@doi [Astronomy Letters]
  {10.1134/S1063773707120043}, 33, 807

\bibitem[\protect\citeauthoryear{Coriat, Fender  \& Dubus}{Coriat
  et~al.}{2012}]{Coriat2012}
Coriat M.,  Fender R.~P.,   Dubus G.,  2012, \mn@doi [MNRAS]
  {10.1111/j.1365-2966.2012.21339.x}, 424, 1991

\bibitem[\protect\citeauthoryear{Corral-Santana, Casares, Mu{\~{n}}oz-Darias,
  Bauer, Mart{\'{i}}nez-Pais  \& Russell}{Corral-Santana
  et~al.}{2016}]{Corral-Santana2016}
Corral-Santana J.~M.,  Casares J.,  Mu{\~{n}}oz-Darias T.,  Bauer F.~E.,
  Mart{\'{i}}nez-Pais I.~G.,   Russell D.~M.,  2016, \mn@doi [A{\&}A]
  {10.1051/0004-6361/201527130}, 587, A61

\bibitem[\protect\citeauthoryear{{De Marco}, Ponti, Mu{\~{n}}oz-Darias  \&
  Nandra}{{De Marco} et~al.}{2015}]{DeMarco2015}
{De Marco} B.,  Ponti G.,  Mu{\~{n}}oz-Darias T.,   Nandra K.,  2015, \mn@doi
  [MNRAS] {10.1093/mnras/stv1990}, 454, 2360

\bibitem[\protect\citeauthoryear{Degenaar \& Wijnands}{Degenaar \&
  Wijnands}{2010}]{Degenaar2010}
Degenaar N.,  Wijnands R.,  2010, \mn@doi [A{\&}A]
  {10.1051/0004-6361/201015322}, 524, A69

\bibitem[\protect\citeauthoryear{Degenaar et~al.,}{Degenaar
  et~al.}{2012}]{Degenaar2012d}
Degenaar N.,  et~al., 2012, \mn@doi [A{\&}A] {10.1051/0004-6361/201118634},
  540, A22

\bibitem[\protect\citeauthoryear{Degenaar, Wijnands  \& Miller}{Degenaar
  et~al.}{2013}]{Degenaar2013}
Degenaar N.,  Wijnands R.,   Miller J.~M.,  2013, \mn@doi [ApJ]
  {10.1088/2041-8205/767/2/L31}, 767, L31

\bibitem[\protect\citeauthoryear{Degenaar, Pinto, Miller, Wijnands, Altamirano,
  Paerels, Fabian  \& Chakrabarty}{Degenaar et~al.}{2017}]{Degenaar2017}
Degenaar N.,  Pinto C.,  Miller J.~M.,  Wijnands R.,  Altamirano D.,  Paerels
  F.,  Fabian A.~C.,   Chakrabarty D.,  2017, \mn@doi [MNRAS]
  {10.1093/mnras/stw2355}, 464, 398

\bibitem[\protect\citeauthoryear{{Del Santo}, Sidoli, Mereghetti, Bazzano,
  Tarana, Ubertini  \& {Del Santo}}{{Del Santo} et~al.}{2007}]{DelSanto2007}
{Del Santo} M.,  Sidoli L.,  Mereghetti S.,  Bazzano A.,  Tarana A.,  Ubertini
  P.,   {Del Santo} M.,  2007, \mn@doi [A{\&}A] {10.1051/0004-6361:20077536},
  468, L17

\bibitem[\protect\citeauthoryear{Falanga, Chenevez, Cumming, Kuulkers, Trap  \&
  Goldwurm}{Falanga et~al.}{2008}]{Falanga2008}
Falanga M.,  Chenevez J.,  Cumming A.,  Kuulkers E.,  Trap G.,   Goldwurm A.,
  2008, \mn@doi [A{\&}A] {10.1051/0004-6361:20078982}, 484, 43

\bibitem[\protect\citeauthoryear{Frank, King  \& Lasota}{Frank
  et~al.}{1987}]{Frank1987}
Frank J.,  King A.~R.,   Lasota J.-P.,  1987, A{\&}A, 178, 137

\bibitem[\protect\citeauthoryear{Garc{\'{i}}a, Steiner, McClintock, Remillard,
  Grinberg  \& Dauser}{Garc{\'{i}}a et~al.}{2015}]{Garcia2015}
Garc{\'{i}}a J.~A.,  Steiner J.~F.,  McClintock J.~E.,  Remillard R.~A.,
  Grinberg V.,   Dauser T.,  2015, \mn@doi [ApJ] {10.1088/0004-637X/813/2/84},
  813, 84

\bibitem[\protect\citeauthoryear{Heil, Vaughan  \& Uttley}{Heil
  et~al.}{2012}]{Heil2012}
Heil L.~M.,  Vaughan S.,   Uttley P.,  2012, \mn@doi [MNRAS]
  {10.1111/j.1365-2966.2012.20824.x}, 422, 2620

\bibitem[\protect\citeauthoryear{{In 't Zand} et~al.,}{{In 't Zand}
  et~al.}{2002}]{IntZand2002}
{In 't Zand} J. J.~M.,  et~al., 2002, \mn@doi [A{\&}A]
  {10.1051/0004-6361:20020631}, 389, 43

\bibitem[\protect\citeauthoryear{{In 't Zand}, Jonker  \& Markwardt}{{In 't
  Zand} et~al.}{2007}]{IntZand2007}
{In 't Zand} J. J.~M.,  Jonker P.~G.,   Markwardt C.~B.,  2007, \mn@doi
  [A{\&}A] {10.1051/0004-6361:20066678}, 465, 953

\bibitem[\protect\citeauthoryear{Jansen et~al.,}{Jansen
  et~al.}{2001}]{Jansen2001}
Jansen F.,  et~al., 2001, \mn@doi [A{\&}A] {10.1051/0004-6361:20000036}, 365,
  L1

\bibitem[\protect\citeauthoryear{Koyama et~al.,}{Koyama
  et~al.}{2007}]{Koyama2007}
Koyama K.,  et~al., 2007, \mn@doi [PASJ] {10.1093/pasj/59.sp1.S23}, 59, S23

\bibitem[\protect\citeauthoryear{Kubota, Tanaka, Makishima, Ueda, Dotani, Inoue
   \& Yamaoka}{Kubota et~al.}{1998}]{Kubota1998}
Kubota A.,  Tanaka Y.,  Makishima K.,  Ueda Y.,  Dotani T.,  Inoue H.,
  Yamaoka K.,  1998, \mn@doi [PASJ] {10.1093/pasj/50.6.667}, 50, 667

\bibitem[\protect\citeauthoryear{Kubota et~al.,}{Kubota
  et~al.}{2004}]{Kubota2004}
Kubota A.,  et~al., 2004, \mn@doi [ApJ] {10.1086/380433}, 601, 428

\bibitem[\protect\citeauthoryear{Kuulkers et~al.,}{Kuulkers
  et~al.}{2013}]{Kuulkers2013}
Kuulkers E.,  et~al., 2013, \mn@doi [A{\&}A] {10.1051/0004-6361/201219447},
  552, A32

\bibitem[\protect\citeauthoryear{Lasota}{Lasota}{2001}]{Lasota2001}
Lasota J.-P.,  2001, \mn@doi [MNRAS] {10.1016/S1387-6473(01)00112-9}, 45, 449

\bibitem[\protect\citeauthoryear{Lin, Remillard  \& Homan}{Lin
  et~al.}{2007}]{Lin2007}
Lin D.,  Remillard R.~A.,   Homan J.,  2007, \mn@doi [ApJ] {10.1086/521181},
  667, 1073

\bibitem[\protect\citeauthoryear{Makishima, Maejima, Mitsuda, Bradt, Remillard,
  Tuohy, Hoshi  \& Nakagawa}{Makishima et~al.}{1986}]{Makishima1986}
Makishima K.,  Maejima Y.,  Mitsuda K.,  Bradt H.~V.,  Remillard R.~A.,  Tuohy
  I.~R.,  Hoshi R.,   Nakagawa M.,  1986, \mn@doi [ApJ] {10.1086/164534}, 308,
  635

\bibitem[\protect\citeauthoryear{Matsuoka \& Asai}{Matsuoka \&
  Asai}{2013}]{Matsuoka2013}
Matsuoka M.,  Asai K.,  2013, \mn@doi [PASJ] {10.1093/pasj/65.2.26}, 65, 26

\bibitem[\protect\citeauthoryear{Miller, Homan  \& Miniutti}{Miller
  et~al.}{2006}]{Miller2006b}
Miller J.~M.,  Homan J.,   Miniutti G.,  2006, \mn@doi [ApJ] {10.1086/510015},
  652, L113

\bibitem[\protect\citeauthoryear{Mitsuda et~al.,}{Mitsuda
  et~al.}{1984}]{Mitsuda1984}
Mitsuda K.,  et~al., 1984, PASJ, 36, 741

\bibitem[\protect\citeauthoryear{Mitsuda et~al.,}{Mitsuda
  et~al.}{2007}]{Mitsuda2007}
Mitsuda K.,  et~al., 2007, \mn@doi [PASJ] {10.1093/pasj/59.sp1.S1}, 59, S1

\bibitem[\protect\citeauthoryear{Miyamoto, Kimura, Kitamoto, Dotani  \&
  Ebisawa}{Miyamoto et~al.}{1991}]{Miyamoto1991}
Miyamoto S.,  Kimura K.,  Kitamoto S.,  Dotani T.,   Ebisawa K.,  1991, \mn@doi
  [ApJ] {10.1086/170837}, 383, 784

\bibitem[\protect\citeauthoryear{Motta, Belloni  \& Homan}{Motta
  et~al.}{2009}]{Motta2009}
Motta S.,  Belloni T.,   Homan J.,  2009, \mn@doi [MNRAS]
  {10.1111/j.1365-2966.2009.15566.x}, 400, 1603

\bibitem[\protect\citeauthoryear{Mu{\~{n}}oz-Darias, Motta  \&
  Belloni}{Mu{\~{n}}oz-Darias et~al.}{2011a}]{Munoz-Darias2011a}
Mu{\~{n}}oz-Darias T.,  Motta S.,   Belloni T.~M.,  2011a, \mn@doi [MNRAS]
  {10.1111/j.1365-2966.2010.17476.x}, 410, 679

\bibitem[\protect\citeauthoryear{Mu{\~{n}}oz-Darias, Motta, Stiele  \&
  Belloni}{Mu{\~{n}}oz-Darias et~al.}{2011b}]{Munoz-Darias2011b}
Mu{\~{n}}oz-Darias T.,  Motta S.,  Stiele H.,   Belloni T.~M.,  2011b, \mn@doi
  [MNRAS] {10.1111/j.1365-2966.2011.18702.x}, 415, 292

\bibitem[\protect\citeauthoryear{Munoz-Darias, Coriat, Plant, Ponti, Fender  \&
  Dunn}{Munoz-Darias et~al.}{2013}]{Munoz-Darias2013}
Munoz-Darias T.,  Coriat M.,  Plant D.~S.,  Ponti G.,  Fender R.~P.,   Dunn R.
  J.~H.,  2013, \mn@doi [MNRAS] {10.1093/mnras/stt546}, 432, 1330

\bibitem[\protect\citeauthoryear{Mu{\~{n}}oz-Darias, Fender, Motta  \&
  Belloni}{Mu{\~{n}}oz-Darias et~al.}{2014}]{Munoz-Darias2014}
Mu{\~{n}}oz-Darias T.,  Fender R.~P.,  Motta S.~E.,   Belloni T.~M.,  2014,
  \mn@doi [MNRAS] {10.1093/mnras/stu1334}, 443, 3270

\bibitem[\protect\citeauthoryear{Nandra, O'Neill, George  \& Reeves}{Nandra
  et~al.}{2007}]{Nandra2007}
Nandra K.,  O'Neill P.~M.,  George I.~M.,   Reeves J.~N.,  2007, \mn@doi
  [MNRAS] {10.1111/j.1365-2966.2007.12331.x}, 382, 194

\bibitem[\protect\citeauthoryear{Nayakshin, Kazanas  \& Kallman}{Nayakshin
  et~al.}{2000}]{Nayakshin2000}
Nayakshin S.,  Kazanas D.,   Kallman T.~R.,  2000, \mn@doi [ApJ]
  {10.1086/309054}, 537, 833

\bibitem[\protect\citeauthoryear{Nelemans \& Jonker}{Nelemans \&
  Jonker}{2010}]{Nelemans2010a}
Nelemans G.,  Jonker P.~G.,  2010, \mn@doi [New Astronomy Reviews]
  {10.1016/j.newar.2010.09.021}, 54, 87

\bibitem[\protect\citeauthoryear{Ng, {D{\'{i}}az Trigo}, {Cadolle Bel}  \&
  Migliari}{Ng et~al.}{2010}]{Ng2010}
Ng C.,  {D{\'{i}}az Trigo} M.,  {Cadolle Bel} M.,   Migliari S.,  2010, \mn@doi
  [A{\&}A] {10.1051/0004-6361/200913575}, 522, A96

\bibitem[\protect\citeauthoryear{Plant, Fender, Ponti, Mu{\~{n}}oz-Darias  \&
  Coriat}{Plant et~al.}{2015}]{Plant2015}
Plant D.~S.,  Fender R.~P.,  Ponti G.,  Mu{\~{n}}oz-Darias T.,   Coriat M.,
  2015, \mn@doi [A{\&}A] {10.1051/0004-6361/201423925}, 573, A120

\bibitem[\protect\citeauthoryear{Ponti, Papadakis, Bianchi, Guainazzi, Matt,
  Uttley  \& Bonilla}{Ponti et~al.}{2012}]{Ponti2012}
Ponti G.,  Papadakis I.,  Bianchi S.,  Guainazzi M.,  Matt G.,  Uttley P.,
  Bonilla N.~F.,  2012, \mn@doi [A{\&}A] {10.1051/0004-6361/201118326}, 542,
  A83

\bibitem[\protect\citeauthoryear{Ponti et~al.,}{Ponti et~al.}{2015}]{Ponti2015}
Ponti G.,  et~al., 2015, \mn@doi [MNRAS] {10.1093/mnras/stv1331}, 453, 172

\bibitem[\protect\citeauthoryear{Ponti, Bianchi, Mu{\~{n}}oz-Darias, De, Fender
   \& Merloni}{Ponti et~al.}{2016}]{Ponti2016}
Ponti G.,  Bianchi S.,  Mu{\~{n}}oz-Darias T.,  De K.,  Fender R.,   Merloni
  A.,  2016, \mn@doi [Astronomische Nachrichten] {10.1002/asna.201612339}, 337,
  512

\bibitem[\protect\citeauthoryear{Ponti et~al.,}{Ponti et~al.}{2017}]{Ponti2017}
Ponti G.,  et~al., 2017, \mn@doi [MNRAS] {10.1093/mnras/stx596}, 468, 2447

\bibitem[\protect\citeauthoryear{Scargle}{Scargle}{1982}]{Scargle1982}
Scargle J.~D.,  1982, \mn@doi [ApJ] {10.1086/160554}, 263, 835

\bibitem[\protect\citeauthoryear{Shidatsu et~al.,}{Shidatsu
  et~al.}{2014}]{Shidatsu2014}
Shidatsu M.,  et~al., 2014, \mn@doi [ApJ] {10.1088/0004-637X/789/2/100}, 789,
  100

\bibitem[\protect\citeauthoryear{Skinner et~al.,}{Skinner
  et~al.}{1987}]{Skinner1987}
Skinner G.~K.,  et~al., 1987, \mn@doi [Nature] {10.1038/330544a0}, 330, 544

\bibitem[\protect\citeauthoryear{Str{\"{u}}der et~al.,}{Str{\"{u}}der
  et~al.}{2001}]{Struder2001}
Str{\"{u}}der L.,  et~al., 2001, \mn@doi [A{\&}A] {10.1051/0004-6361:20000066},
  365, L18

\bibitem[\protect\citeauthoryear{Takahashi et~al.,}{Takahashi
  et~al.}{2007}]{Takahashi2007}
Takahashi T.,  et~al., 2007, \mn@doi [PASJ] {10.1093/pasj/59.sp1.S35}, 59, S35

\bibitem[\protect\citeauthoryear{Tetarenko, Sivakoff, Heinke  \&
  Gladstone}{Tetarenko et~al.}{2016}]{Tetarenko2016}
Tetarenko B.~E.,  Sivakoff G.~R.,  Heinke C.~O.,   Gladstone J.~C.,  2016,
  \mn@doi [ApJS] {10.3847/0067-0049/222/2/15}, 222, 15

\bibitem[\protect\citeauthoryear{Tomsick, Walter, Kaaret, Rodriguez  \&
  Chaty}{Tomsick et~al.}{2007}]{Tomsick2007}
Tomsick J.~A.,  Walter R.,  Kaaret P.,  Rodriguez J.,   Chaty S.,  2007, The
  Astronomer's Telegram, No.1189, 1189

\bibitem[\protect\citeauthoryear{Turner et~al.,}{Turner
  et~al.}{2001}]{Turner2001}
Turner M.,  et~al., 2001, \mn@doi [A{\&}A] {10.1051/0004-6361:20000087}, 365,
  L27

\bibitem[\protect\citeauthoryear{Vaughan, Edelson, Warwick  \& Uttley}{Vaughan
  et~al.}{2003}]{Vaughan2003}
Vaughan S.,  Edelson R.,  Warwick R.~S.,   Uttley P.,  2003, \mn@doi [MNRAS]
  {10.1046/j.1365-2966.2003.07042.x}, 345, 1271

\bibitem[\protect\citeauthoryear{Verner, Ferland, Korista  \& Yakovlev}{Verner
  et~al.}{1996}]{Verner1996}
Verner D.,  Ferland G.,  Korista K.,   Yakovlev D.,  1996, \mn@doi [ApJ]
  {10.1086/177435}, 465, 487

\bibitem[\protect\citeauthoryear{White \& Mason}{White \&
  Mason}{1985}]{White1985}
White N.~E.,  Mason K.~O.,  1985, \mn@doi [Space Sci. Rev.]
  {10.1007/BF00212883}, 40, 167

\bibitem[\protect\citeauthoryear{Wijnands, Degenaar, {Armas Padilla},
  Altamirano, Cavecchi, Linares, Bahramian  \& Heinke}{Wijnands
  et~al.}{2015}]{Wijnands2015}
Wijnands R.,  Degenaar N.,  {Armas Padilla} M.,  Altamirano D.,  Cavecchi Y.,
  Linares M.,  Bahramian A.,   Heinke C.~O.,  2015, \mn@doi [MNRAS]
  {10.1093/mnras/stv1974}, 454, 1371

\bibitem[\protect\citeauthoryear{Wilms, Allen  \& McCray}{Wilms
  et~al.}{2000}]{Wilms2000}
Wilms J.,  Allen A.,   McCray R.,  2000, \mn@doi [ApJ] {10.1086/317016}, 542,
  914

\bibitem[\protect\citeauthoryear{Yamada et~al.,}{Yamada
  et~al.}{2012}]{Yamada2012}
Yamada S.,  et~al., 2012, \mn@doi [PASJ] {10.1093/pasj/64.3.53}, 64, 53

\bibitem[\protect\citeauthoryear{Zdziarski, Johnson  \& Magdziarz}{Zdziarski
  et~al.}{1996}]{Zdziarski1996}
Zdziarski A.,  Johnson W.,   Magdziarz P.,  1996, MNRAS, 283, 193

\bibitem[\protect\citeauthoryear{Zycki, Done  \& Smith}{Zycki
  et~al.}{1999}]{Zycki1999}
Zycki P.~T.,  Done C.,   Smith D.~A.,  1999, \mn@doi [MNRAS]
  {10.1046/j.1365-8711.1999.02885.x}, 309, 561

\bibitem[\protect\citeauthoryear{{in 't Zand}, Cornelisse  \& M{\'{e}}ndez}{{in
  't Zand} et~al.}{2005}]{IntZand2005}
{in 't Zand} J. J.~M.,  Cornelisse R.,   M{\'{e}}ndez M.,  2005, \mn@doi
  [A{\&}A] {10.1051/0004-6361:20052955}, 440, 287

\bibitem[\protect\citeauthoryear{van~der Klis}{van~der
  Klis}{2006}]{VanderKlis2006}
van~der Klis M.,  2006, In: Compact stellar X-ray sources. Edited by Walter
  Lewin {\&} Michiel van der Klis. Cambridge Astrophysics Series

\makeatother
\end{thebibliography}



%
%


\bsp	
\label{lastpage}
\end{document}